\documentclass{amsart}

\usepackage{epsfig,amsmath,amssymb}

\newcommand{\half}{\frac{1}{2}}
\newcommand{\re}{\mbox{Re}}
\newcommand{\om}{\omega}
\def\l{\lambda}

\def\sgn{\mathop{\rm sgn}\nolimits}

\newcommand{\C}{{\mathbf C}}

\newcommand{\vctr}[2]{\left(\begin{array}{c} {#1} \\ {#2}\end{array}\right)}
\newcommand{\mat}[4]{\left(\begin{array}{cc} {#1} & {#2} \\ {#3} & {#4}
\end{array}\right)}

\newtheorem{proposition}{Proposition}

\begin{document}

\title[Symbolic analysis of PWL maps]{Symbolic analysis for some
planar piecewise linear maps}

\author{Xin-Chu Fu \and
Peter Ashwin}


\address{
 Xin-Chu Fu, Department of Mathematics,
 Shanghai University,
 Shanghai 200436, P. R. CHINA
}

\email{x.c.fu@163.com}

\address{
Peter Ashwin, School of Mathematical Sciences,
University of Exeter,
Exeter EX4 4QE, U. K.
}

\email{P.Ashwin@ex.ac.uk}

\date{\today}

\begin{abstract}
In this paper a class of linear maps on the 2-torus and some planar
piecewise isometries are discussed. For these discontinuous maps, by
introducing codings underlying
the map operations, symbolic descriptions of the dynamics and admissibility
conditions for itineraries are given, and explicit expressions in terms of
the codings for periodic points are presented.
\end{abstract}


\maketitle
\section{Introduction}
\label{sec_int}

In contrast to continuous dynamical systems, there is not yet a
systematic theory and only limited effective methods available in
the study of dynamical systems with discontinuities, but symbolic
dynamics may be a useful tool in this study. For example, in
\cite{ChuaLin90} a symbolic dynamics approach for the study of
fractal pattern in second-order non-linear digital filters is
presented; in \cite{Zheng92, ZH94,Hao89,Hao_Zheng98} symbolic
dynamics for some invertible 2-dimensional discontinuous
hyperbolic maps is discussed; in \cite{Goetz99} symbolic dynamics
analysis for generalized piecewise isometries is given, and some
interesting examples and open questions are presented; in
\cite{Kopf00} symbolic analysis for piecewise continuous and
piecewise monotone transformations on the interval $I=[0,1]$ is
discussed, and by giving necessary and sufficient condition for an
arbitrary symbol sequence to be an $n$-address, an algorithm to
calculate the topological entropy is given. Most recently, in
\cite{AdlKitTre99} some orbits of a class of non-ergodic piecewise
affine maps of the 2-torus are described in terms of a symbol
shift obtained from  the ``triadic odometer'' substitution rule.

In this paper we show that symbolic dynamics method is helpful
when considering a class of piecewise linear maps on the 2-torus,
i.e., the maps $(x',y')=f(x,y)$ on $X=[0,1)^2$ with the form
below:
\begin{equation}\label{eq_torusmap}
\left\{\begin{array}{l}
x'=ax+by ~(\bmod ~1)\\
y'=cx+dy ~(\bmod ~1)
\end{array}\right.
\end{equation}
where $a,b,c,d \in {\mathbb R}$, and this map can be thought of as
a map $f=g\circ M$ where
$$
M=\mat{a}{b}{c}{d}
$$
(we also write $M=(a,b;c,d)$ for convenience) and $g(x)=x-\lfloor
x\rfloor$ is a map that takes modulo $1$ in each component, and we
assume the determinant $\det M\ne 0$. As pointed out by Adler in
\cite{Adler98}, the symbolic dynamics even for hyperbolic
automorphisms of the 2-torus (i.e., all coefficients in
(\ref{eq_torusmap}) are integers, and the map is hyperbolic type)
remains a fertile area for research, while there are some results
in higher dimensions. In general there are still few general
methods for the class of maps (\ref{eq_torusmap}), the symbolic
analysis in this paper represent some technical steps towards this.
We focus on the symbolic dynamics for this class of maps,
and we present some necessary and sufficient conditions for admissibility
of sequences. We also give some results concerning the measure of
the maximal invariant set of certain non-invertible maps of this
type generalizing a result of \cite{AshFuNisZyc}.

In \cite{AshwinFu01, AFD01} for planar piecewise isometries we
have introduced symbolic codings underlying map operations. It has
been shown that this kind of codings is helpful for revealing the
dynamical properties of the maps. In this paper we use this idea
to symbolically analyze the maps of the form (\ref{eq_torusmap}).

In \cite{AshFuNisZyc} it is shown that for the parabolic map of
the form (\ref{eq_torusmap}) (i.e., $\det M =1$, and the trace
$a+d=2$) the 2-dimensional systems have 1-dimensional dynamical
characteristics, that is, the system possesses invariant straight
line segments along the eigen-direction. In \cite{Zheng92} it is
shown that the T\'el map, a piecewise linear hyperbolic map
defined on the whole plane, can be decomposed, along stable and
unstable manifolds, into two coupled one-dimensional maps. Based
on these ideas we perform a dimension reduction in certain cases
of the general 2-torus maps (\ref{eq_torusmap}).

In \cite{Zheng92} for the T\'el map, the orbit of a given point is
encoded according to the sign of $x$ in the $n$-th image and the
sign of $y$ in the $m$-th pre-image, and the symbolic sequences
define an ordering of the contracting and expanding foliations of
the phase plane. Then admissibility conditions for symbolic
sequences can be given by ordering rules of the foliations. In
this paper we generalize these ideas by introducing codings
underlying the map operations, we give symbolic descriptions of
the dynamics and admissibility conditions for itineraries, and
present explicit expressions in terms of the codings for periodic
points.

\section[Partition and coding for linear maps]
{Partition and coding for linear maps on the $2$-torus}
\label{sec_linear}

The maps (\ref{eq_torusmap}) can be classified into three types,
depending on the eigenvalues $\lambda_{1,2}$ of $M$ we refer to
the map as type I ($\lambda_2>\lambda_1$), type II
($\lambda_1=\lambda_2$) or type III
($\lambda_1=\overline{\lambda}_2\neq \lambda_2$). When $\det M=1$,
i.e., the map is area-preserving, then we can refer to a type I,
II or III map as being hyperbolic, parabolic or elliptic,
respectively, as we discussed in \cite{AshFuNisZyc}.

\begin{figure}
\begin{center}
\mbox{\epsfig{file=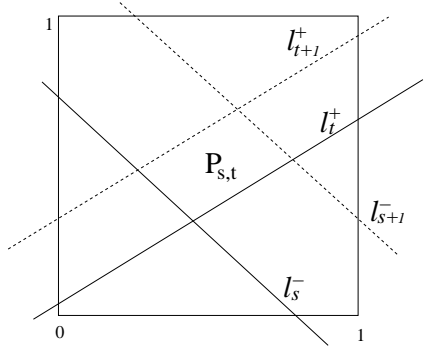,width=6.0cm}}
\end{center}
\caption[The partition for the linear torus map]
{\label{fig_parti} \em The partition for the torus map
(\ref{eq_torusmap}), where $l^-_n$ is the straight line $ax+by=n$
and $l^+_m$ is the straight line $cx+dy=m$.}
\end{figure}

We code the orbits of the torus map (\ref{eq_torusmap}) by 2-d
vector $(s,t)$ based on the following partition of the phase space
(See figure~\ref{fig_parti})
\begin{eqnarray*}
[0,1)^2 &=& \bigcup_{s,t}P_{s,t},\ S\le s\le S',\ T\le t\le T', \\
P_{s,t} &=& \{(x,y)\in[0,1)^2, s\le ax+by<s+1, \ t\le cx+dy<t+1\},
\end{eqnarray*}
where $T=\min\{\lfloor ax+by\rfloor\}, \ T'=\max\{\lfloor
ax+by\rfloor \},\ S=\min\{\lfloor cx+dy\rfloor\}, \
S'=\max\{\lfloor cx+dy\rfloor \}.$

This coding underlies the map operations, as the system
(\ref{eq_torusmap}) can be rewritten as
\begin{equation}\label{eq_coding}
\left\{\begin{array}{l}
x'=ax+by-s\\
y'=cx+dy-t
\end{array}. \right.
\end{equation}

For all $(x,y)\in [0,1)^2$ we can define a unique itinerary, this
is a map $\iota: [0,1)^2 \rightarrow \Sigma(N)$ with $\Sigma(N)$
the set of infinite words with $N$ letters, i.e., the $N$ pairs
$(s,t)$, where $N=(T'-T+1)(S'-S+1)$,
$$
\iota(x,y)={\bf r}=(r_0r_1\cdots), \mbox{ where } r_j=(s_j, t_j)
\mbox{ if } f^j(x,y)\in P_{s_j, t_j}.
$$
Recall that a symbolic sequence ${\bf r}\in \Sigma(N)$ is
admissible if there exists a point $(x,y) \in [0,1)^2$ such that
$\iota(x,y)={\bf r}$. Let $\Sigma_f\subseteq \Sigma(N)$ be the
subset of all admissible sequences, then the following diagram
commutes:
$$
\begin{array}{ccc}
 [0,1)^2 & \mathop{\longrightarrow}\limits^{f} & [0,1)^2 \\
\iota \downarrow &  & \downarrow \iota \\
\Sigma_f & \mathop{\longrightarrow}\limits^{\sigma} & \Sigma_f
\end{array}
$$
So characterizing the structure of $\Sigma_f$ would be helpful to
the understanding of the dynamics of $f$ on $[0,1)^2$.

\section{Type I and type II linear torus maps}
\label{sec_hyper_para}

We discuss in this section the type I and II cases of
(\ref{eq_torusmap}), namely, the cases where the eigenvalues are
real.

\begin{figure}
\begin{center}
\mbox{\epsfig{file=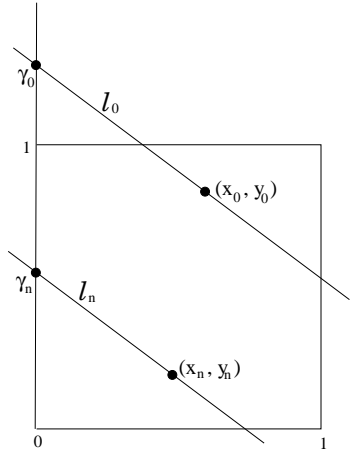,width=5.0cm}}
\end{center}
\caption[The $n$-th iteration of the linear torus map]
{\label{fig_eigendir} \em The $n$-th iteration $(x_n, y_n)$ with
coding $(s_n, t_n)$ is located on the straight line $l_n: \
y=\frac{1}{\tau}x+\gamma_n$.}
\end{figure}

Suppose the $n$-th iteration $(x_n, y_n)$ of the map with coding
$(s_n, t_n)$ is located on the straight line $l_n$:
$$
y=\frac{1}{\tau}x+\gamma_n,
$$
where $n\ge 0$, and $(\tau, 1)$ is an eigenvector for
$M=(a,b;c,d)$. Then we have:

\begin{proposition}\label{prop1-d}
A type I or type II map (\ref{eq_torusmap}) sends points on line
$l_n$ to points on line $l_{n+1}$, where
$\gamma_{n+1}=\frac{1}{\tau}s_n-t_n+\frac{d\tau-b}{\tau}\gamma_n,
\ n\ge 0$. In this sense, the $2$-dimensional system
(\ref{eq_torusmap}) possesses $1$-dimensional dynamics.
\end{proposition}

\proof
We have
\begin{eqnarray}\label{eqa} x_{n+1} &=& ax_n+by_n-s_n\\
\label{eqb} y_{n+1} &=&
cx_n+dy_n-t_n\\ \label{eqc} y_n &=& \frac{1}{\tau}x_n+\gamma_n\\
\label{eqd} y_{n+1} &=& \frac{1}{\tau}x_{n+1}+\gamma_{n+1}
\end{eqnarray}
Substituting (\ref{eqa}) and (\ref{eqb}) into (\ref{eqd}), we
have:
\begin{equation}\label{eqe}
cx_n+dy_n-t_n=\frac{1}{\tau}(ax_n+by_n-s_n)+\tau_{n+1},
\end{equation}
and substituting (\ref{eqc}) into (\ref{eqe}), we get:
\begin{equation}\label{eq_gamma}
\gamma_{n+1}=\frac{1}{\tau}s_n-t_n+\frac{d\tau-b}{\tau}\gamma_n, \
n\ge 0
\end{equation}
That is, a type I or type II map (\ref{eq_torusmap}) sends points
on line $l_n$ to points on line $l_{n+1}$, and line $l_{n+1}$ can
be determined through (\ref{eq_gamma}) by $l_n$ and coding
$(s_n,t_n)$. If we identify all points on line $l_i$ to be a
single point, $i\geq 0$, then there is a factor system that is
$1$-dimensional.
\qed

Rewrite (\ref{eq_gamma}) as
\begin{equation*}
\gamma_n=-\frac{1}{\tau}\frac{\tau}{d\tau-b}s_n+\frac{\tau}{d\tau-b}t_n
+\frac{\tau}{d\tau-b}\gamma_{n+1},
\end{equation*}
so we get the relation among $\gamma_j$ and the codings
$(s_j,t_j)$ for $0\le n\le j \le n+J+1$:
\begin{equation}\label{eq_gamma2}
\gamma_n=-\frac{1}{\tau}\sum_{j=0}^{J}
\left(\frac{\tau}{d\tau-b}\right)^{j+1}s_{n+j}+
\sum_{j=0}^{J}\left(\frac{\tau}{d\tau-b}\right)^{j+1}t_{n+j}
+\left(\frac{\tau}{d\tau-b}\right)^{J+1}\gamma_{n+J+1}
\end{equation}
Under the condition that $|\frac{\tau}{d\tau -b}|<1$, the last
term in (\ref{eq_gamma2}) vanishes as $J$ goes to infinity, and we
have:
\begin{equation}\label{eqf}
\gamma_n=-\frac{1}{\tau}\sum_{j\ge 0}
\left(\frac{\tau}{d\tau-b}\right)^{j+1}s_{n+j}+ \sum_{j\ge
0}\left(\frac{\tau}{d\tau-b}\right)^{j+1}t_{n+j}
\end{equation}

Whenever ${\bf r}$ is a periodic sequence, that is,
$$
{\bf r}=(r_0r_1\cdots)=P(r_0r_1\cdots r_{n-1}),
$$
where $P$ denotes the periodic concatenation of its argument, and
if this is admissible then there will be some points $(x,y)$ with
$\iota(x,y)={\bf r}$ and $\{\gamma_j, \ j\ge 0\}$ calculated by
(\ref{eqf}) is also a periodic sequence, and
$$
(\gamma_0\gamma_1\cdots \gamma_j\cdots)=P(\gamma_0\gamma_1\cdots
\gamma_{n-1}).
$$
Therefore, if the itinerary of $(x_0, y_0)$ is periodic, its orbit
$\{(x_j,y_j), \ j\ge 0\}$ will jump periodically among $n$
straight line segments $l_j: \ y=\frac{1}{\tau}x+\gamma_j, \ 0\le
j \le n-1$ with the same slope. This is similar to quasi-periodic
motion in elliptic systems in which quasi-periodic points jump
periodically among finite number of circles with the same radius.
For system (\ref{eq_torusmap}) in type I and type II cases we call
an orbit starting from $(x_0, y_0)$ periodically coded if
$\iota(x_0,y_0)$ is a periodic sequence.

\begin{figure}
\begin{center}
\mbox{\psfig{file=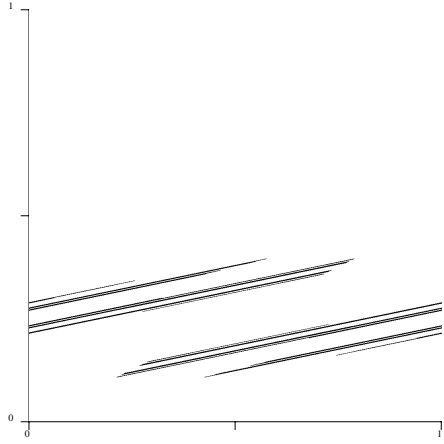,width=6.0cm,bbllx=0.7cm,bblly=4.2cm,
bburx=20cm,bbury=23cm,clip=}}
\end{center}
\caption[Numerical result for a linear hyperbolic map]
{\label{fig_finite_lines} \em Numerical result for hyperbolic
system with $a=\frac{3}{2}, b=1,c=\frac{1}{4}, d=\half$. For big
$n$, $\gamma_n$ takes a small subset of its possible values, the
whole phase space is shrunk into a Cantor set consisting of
uncountable parallel
straight line segments.}
\end{figure}

Numerical results show that for some parameters, asymptotically,
$\gamma_n$ takes values within a quite narrow range, that means
under the dynamics the whole phase space can be shrunk into a very
small subset. (See Figure~\ref{fig_finite_lines}).

In some cases, $\gamma_n$ may only take a finite number of values
for each orbit. For example, we have

\begin{proposition}
When $M=(1+A, -A; A, 1-A), A $ is any real number, then $\gamma_n$
takes at most two values for each orbit.
\end{proposition}

\proof This can be shown as follows. Now $f$ is an area preserving
parabolic map with eigenvalue $\lambda=1$. So $\tau=1$ and
therefore $\frac{d\tau-b}{\tau}=1$. According to (\ref{eq_gamma}),
$$
\gamma_{n+1}=s_n-t_n+\gamma_n, \ n\ge 0,
$$
so $\gamma_{n+1}-\gamma_n, \ n\ge 0$ are all integers. But for
$\tau=1, \ -1< \gamma_n <1$, therefore $\gamma_n$ takes at most
two values:
$$
\gamma_n=\left\{
\begin{array}{l}
\gamma_0 \\
\gamma_0-\sgn \gamma_0
\end{array}. \right.
$$
When $\gamma_0=0$, then $\gamma_n=0$ for all $n$. In fact, all
points on the line $y=x$ are fixed points of $f$. For $x_0\ne
y_0$, the motion of the orbit from the initial point $(x_0,y_0)$
will just restrict on two lines $l: \ y=x+\gamma_0$ and $l': \
y=x+\gamma_0-\sgn \gamma_0$, where $\gamma_0=y_0-x_0$. \qed

For a general area preserving torus map $f$ discussed in
\cite{AshFuNisZyc}  , we have $M=M_{A,\alpha}=(1+A, \alpha^{-1}A;
-\alpha A, 1-A)$. Then $\tau=-\alpha^{-1}$ as the eigenvalue
$\lambda=1$, and therefore
$$
\frac{d\tau-b}{\tau}=\frac{-(1-A)\alpha^{-1}-\alpha^{-1}A}{-\alpha^{-1}}=1.
$$
According to (\ref{eq_gamma}),
$$
\gamma_{n+1}=-\alpha s_n-t_n+\gamma_n,~~n\ge 0.
$$
so we have
\begin{equation}\label{eqgamma}
\gamma_{n+1}-\gamma_n=-\alpha s_n-t_n,~~n\ge 0.
\end{equation}

\begin{proposition}  \label{propgamma}
When $M=M_{A,\alpha}$, and $\alpha$ is rational, then $\gamma_n$
takes a finite number of values for each orbit.
\end{proposition}

\proof  Now $f$ is an area-preserving semi-rational parabolic map.
Since $\alpha$ is rational, from (\ref{eqgamma}) there is a
integer $N$ such that $N(\gamma_{n+1}-\gamma_n), n\ge 0$ are all
integers. Note that $\gamma_n, n\ge 0$ are bounded, so $\gamma_n$
can only take a finite number of values for each orbit. \qed

In \cite{AshFuNisZyc} it is shown that for a rational parabolic
torus map ($A, \alpha$ are both rational) the maximal invariant
set $X^+$ has positive Lebesgue measure. Here we further have

\begin{proposition}
For a semi-rational parabolic torus map ($\alpha$ is rational) the
maximal invariant set $X^+$ has positive Lebesgue measure.
\end{proposition}

\proof  {}From Proposition~\ref{propgamma} an orbit of $f$ is
contained in finite number of straight lines $l_n: y=-\alpha
x+\gamma_n.$  Note that $f$ is piecewise continuous and preserves
Lebesgue measure locally. So the intersection of the maximal
invariant set $X^+$ with these lines contains at least one line
segment $L$ with positive length (in fact, the intersection
usually contains countable line segments possessing positive
length). By similar argument used in this paper, the location of
$L$ can be determined {\em linearly} by the codings of the orbit,
$\gamma_n$ and $\alpha$. Let $\gamma_0$ vary continuously in a
certain small range such that the codings keep unchanged and
(\ref{eqgamma}) keep valid, then the location of $L$ and its
length changes continuously. In this way we get a small
2-dimensional piece with positive Lebesgue measure. Because this
piece is contained in $X^+$,  $X^+$ therefore has positive
Lebesgue measure. \qed

The above proposition partially solved an open problem in
\cite{AshFuNisZyc}. It still remains unknown whether $X^+$ has
positive measure for irrational parabolic torus maps.

\subsection{The cases for $c=1, d=0$}
\label{sec_c1d0}

We can check that we can take $\tau=(\l-d)/c$ where $\l$ is an
eigenvalue of $M$. When $c=1, d=0$, then $\tau=\l$, and the system
(\ref{eq_torusmap}) becomes
\begin{equation}\label{eq_torusmap2}
\left\{ \begin{array}{l}
x'=ax+by ~(\bmod ~1)\\
y'=x
\end{array}\right.
\end{equation}

We suppose $a,b>0$. When $a+b\le 1$, the map (\ref{eq_torusmap2})
is continuous, so we further suppose $a+b>1$. Therefore system
(\ref{eq_torusmap2}) is type I with two eigenvalues $\l_{1,2}$,
and $-b<\l_1<0,\ \l_2 > 1$.

We can code the orbits of the torus map (\ref{eq_torusmap2}) by
just a number $s$, rather than by a $2$-d vector $(s,t)$, based on
the following partition of the phase space
\begin{eqnarray*}
[0,1)^2 &=& \bigcup_s P_s,\ s\in \{0,1,\cdots,[a+b]\}\\
P_s &=& \{(x,y)\in[0,1)^2, s\le ax+by<s+1\},
\end{eqnarray*}

Suppose an orbit from $(x_0,y_0)\in [0,1)^2$ with the symbolic
sequence $(s_0s_1\cdots s_j \cdots)$ as its itinerary, and suppose
the $n$-th iteration $(x_n,y_n)$ is on the straight line $l_n$
$$
y=\frac{1}{\l_1}x+\alpha_n, \ n\ge 0,
$$
which relates to the stable manifold direction $(\lambda_1,1)$,
then from (\ref{eqf}) we have
\begin{equation}\label{eqg}
\alpha_n=\frac{1}{b}\sum_{j\ge 0}\frac{s_{n+j}}{\l_2^j}.
\end{equation}
Here the infinite series does converge as $\l_2>1$.

\begin{proposition}\label{propadmiss}
The admissibility condition for symbolic sequences $(s_0s_1\cdots
s_j \cdots)$ is
$$
\sum_{j\ge 0}\frac{s_j}{\l_2^j}<b-\frac{b}{\l_1}=b+\l_2;
$$
and the condition for a symbolic sequence $(s_0s_1\cdots s_j
\cdots)$ to be the itinerary of a given point $(x_0,y_0)$ is
$$
\sum_{j\ge 0}\frac{s_j}{\l_2^j}=by_0+\l_2 x_0.
$$
\end{proposition}

\proof
For an admissible symbolic sequence $(s_0s_1\cdots s_j
\cdots)$, a point $(x_n,y_n)$ with coding $s_n$ on the straight
line segment
$$
y=\frac{1}{\l_1}x+\frac{1}{b}\sum_{j\ge 0}\frac{s_{n+j}}{\l_2^j}
$$
is mapped to a point $(x_{n+1},y_{n+1})$ with coding $s_{n+1}$ on
the straight line segment
$$
y=\frac{1}{\l_1}x+\frac{1}{b}\sum_{j\ge
0}\frac{s_{n+j+1}}{\l_2^j}.
$$
So whenever an initial point $(x_0,y_0)$ is on the straight line
segment $l_0$, then $(x_n,y_n)$ would be on the straight line
segment $l_n$. This process can continue if $l_0$ intersects with
$[0,1)^2$, so we have one of the admissibility conditions for
symbolic sequences as follows
\begin{equation}\label{eq_admis}
\sum_{j\ge 0}\frac{s_j}{\l_2^j}<b-\frac{b}{\l_1}=b+\l_2.
\end{equation}

For a given point $(x_0,y_0)\in [0,1)^2$, let its itinerary be
$$
\iota(x_0,y_0)=(s_0s_1\cdots s_j \cdots).
$$
Suppose $y_0=\frac{1}{\l_1}x_0+K$, then $K=y_0-\frac{1}{\l_1}x_0$,
so we should have $\alpha_0=K$, so we have the conditions for a
symbolic sequence to be the itinerary of the given point
$(x_0,y_0)$ as follows:
\begin{equation}\label{eq_admis2}
\sum_{j\ge 0}\frac{s_j}{\l_2^j}=by_0+\l_2 x_0.
\end{equation}
\qed

The set of points $C({\bf s})=\iota^{-1}({\bf s})$ with the same
itinerary ${\bf s}$ we refer to as a {\em cell}.

For an admissible symbolic sequence ${\bf s}=(s_0s_1\cdots s_j
\cdots)$, the cell $C({\bf s})$ is the union of straight line
segments.

For a fixed point of $f$ defined by (\ref{eq_torusmap2}), its
itinerary is also a fixed point for the shift map. We can list all
admissible symbolic sequences for fixed points as
$$
{\bf s_j}=(jj\cdots j \cdots), \ \ 0\le j \le [a+b]-2,
$$
then
$$
\alpha_n({\bf s_j})=\frac{1}{b}\sum\limits_{i\ge
0}\frac{j}{\l_2^i}
=\frac{j}{b}\frac{\lambda_2}{\lambda_2-1}=\alpha(j),
$$
therefore a fixed point $(x_j, y_j)$ should satisfy
\begin{equation*}
\left\{ \begin{array}{l}
y_j=\frac{1}{\l_1}x_j+\alpha(j)\\
y_j=x_j
\end{array}, \right.
\end{equation*}
solve this we get
$$
x_j=y_j= \frac{j}{b}\frac{\lambda_1 \lambda_2}{\lambda_1
\lambda_2- (\lambda_1+\lambda_2)+1}=\frac{j}{a+b-1}.
$$
So all fixed points of $f$ are
$$
F_j=(\frac{j}{a+b-1},\frac{j}{a+b-1}), \ \ j=0,1,\cdots,[a+b]-2.
$$

The above discussion show that there is a one-to-one
correspondence between admissible fixed sequences and fixed points
of $f$. Fixed points are period-1 points. For general periodic
points, we have

\begin{proposition}
There is a one to one correspondence between admissible periodic
symbolic sequences and periodic points of $f$ defined by
(\ref{eq_torusmap2}).
\end{proposition}

\proof To see this, suppose ${\bf s}$ is an admissible
$n$-periodic sequence,
$$
{\bf s}=(s_0s_1\cdots s_j\cdots)=P(s_0s_1\cdots s_{n-1}),
$$
and there exists a point $(x_0,y_0)\in \iota^{-1}({\bf s})$ that
is a $n$-periodic point of $f$. From $s_{j+n}=s_j$ we have
\begin{equation}\label{eq_periodic_alpha}
\alpha_i=\frac{1}{b}\frac{\lambda^n_2}{\lambda^n_2-1}
\sum_{j=0}^{n-1}\frac{s_{j+i}}{\lambda_2^j}, \ i\ge 0,
\end{equation}
therefore $(\alpha_0\alpha_1\cdots \alpha_i \cdots)=
P(\alpha_0\alpha_1\cdots \alpha_{n-1})$. Use the facts that
$y_i=x_{i-1}, \ i\ge 1, \ y_0=y_n=x_{n-1}$, we have
\begin{equation*}
\left\{ \begin{array}{l}
x_i=\frac{1}{\l_1}x_{i+1}+\alpha_{i+1}\\
x_{i+n}=x_i
\end{array}, \ i\ge 0, \right.
\end{equation*}
solve this we obtain
\begin{equation}\label{eq_periodic_xiyi}
\left\{ \begin{array}{l}
x_i=\frac{\lambda^n_1}{\lambda^n_1-1}\sum\limits_{k=0}^{n-1}
\frac{\alpha_{i+1+k}}{\lambda_1^k}\\
y_i=\frac{\lambda^n_1}{\lambda^n_1-1}\sum\limits_{k=0}^{n-1}
\frac{\alpha_{i+k}}{\lambda_1^k}
\end{array}, \ i\ge 0 \right.
\end{equation}
We can check that $\{(x_i,y_i),i\ge 0\}$ determined by
(\ref{eq_periodic_xiyi}) is a $n$-periodic orbit. And from
(\ref{eq_periodic_alpha}) and (\ref{eq_periodic_xiyi}), the
periodic point $(x_0,y_0)$ in $\iota^{-1}({\bf s})$ is uniquely
determined by the codings $s_0, s_1, \cdots, s_{n-1}$. \qed

\subsection{The cases for $a=b=1$}
\label{sec_a1b1}

Below we discuss a more specific case, $a=b=1$. Now the map $f$ is
invertible, and is continuous on the 2-torus topology. In this
case Adler's method in \cite{Adler98} can be applied to discuss
its symbolic dynamics.

The method in \cite{Adler98} is for hyperbolic linear torus
automorphism, that is, the map is hyperbolic and the entries of
the matrix $M$ are integers, and so the map is invertible and
continuous. We can also apply our method described in
section~\ref{sec_hyper_para} to give a symbolic description for
the dynamics of the torus automorphism. Moreover, the method
discussed below can be also applicable to all invertible
hyperbolic linear torus maps of the form (\ref{eq_torusmap}) which
are not necessarily continuous.

For $(x,y)\in [0,1)^2$, there is a unique bi-infinite symbolic
sequence
$$
{\bf s}= (\cdots s_{-j}\cdots s_{-1}\cdot s_0s_1\cdots s_j \cdots)
$$
such that
$$
{\bf s}=\iota(x,y) \ \mbox{ if } f^j\in P_{s_j}, \ j\in {\mathbb
Z}.
$$

The two eigenvalues of $M$ are $\lambda_{1,2}=\half (1\pm
\sqrt{5})$, and ${\bf v}_{1,2}=(\lambda_{1,2}, 1)$ are two
corresponding eigenvectors.

\begin{proposition}\label{propadmiss2}
The admissibility conditions for a symbolic sequence $(\cdots
s_{-1}\cdot s_0s_1\cdots)$ are
$$
\left\{ \begin{array}{rl} 0\le & \sum\limits_{j\le
-1}(-1)^j\lambda_2^js_j+
\sum\limits_{j\ge 0}\lambda_2^{-j}s_j < \lambda_2-\lambda_1 \\
0\le &  \sum\limits_{j\ge 0}\lambda_2^{-j}s_j-
\lambda_2^2\sum\limits_{j\le -1}(-1)^j\lambda_2^{j}s_j < 1+
\lambda_2^2
\end{array};\right.
$$
and the conditions for a symbolic sequence $(\cdots s_{-1}\cdot
s_0s_1\cdots)$ to be the itinerary of a given point $(x_0,y_0)$
are
$$
\left\{ \begin{array}{l}
\sum\limits_{j\ge 0}\lambda_2^{-j}s_j=y_0-\frac{1}{\l_1}x_0\\
\sum\limits_{j\le -1}\lambda_2^{j}s_j=\frac{1}{\l_2}x_0-y_0
\end{array}.\right.
$$
And there is a one-to-one correspondence between all points in the
phase space (and the orbits from them) and all admissible symbolic
sequences (and their shifts).
\end{proposition}

\begin{figure}
\begin{center}
\mbox{\epsfig{file=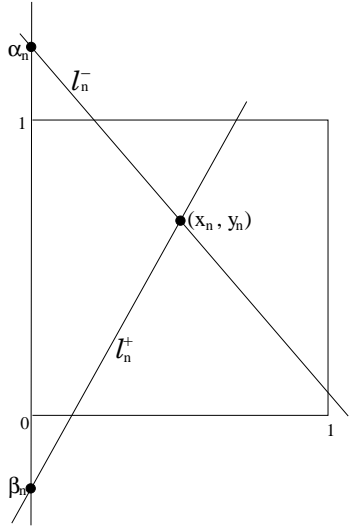,width=5.0cm}}
\end{center}
\caption[The stable and unstable manifolds] {\label{fig_manifolds}
\em The stable and unstable manifolds $l^-_n: \
y=\frac{1}{\lambda_1}x +\alpha_n$ and $l^+_n: \
y=\frac{1}{\lambda_2}x +\beta_n$ intersect at $(x_n,
y_n)=f^n(x_0,y_0)$.}
\end{figure}

\proof Suppose the $n$-th iteration $(x_n, y_n)$ is the
intersection of two straight lines $l^-_n: \
y=\frac{1}{\lambda_1}x +\alpha_n$ and $l^+_n: \
y=\frac{1}{\lambda_2}x +\beta_n$ (See figure~\ref{fig_manifolds}).
Then from (\ref{eqg}) we have
\begin{equation}\label{eqga}
\alpha_n=\sum_{j\ge 0}\frac{s_{n+j}}{\l_2^j}.
\end{equation}

Similar to (\ref{eqa})-(\ref{eqd}) we have
\begin{eqnarray}\label{eqa2}
x_{n} &=& x_{n-1}+y_{n-1}-s_{n-1}\\ \label{eqb2} y_{n} &=& x_{n-1}
\\ \label{eqc2} y_{n-1} &=&
\frac{1}{\lambda_2}x_{n-1}+\beta_{n-1}\\ \label{eqd2} y_{n} &=&
\frac{1}{\lambda_2}x_{n}+\lambda_{n}
\end{eqnarray}
Substitute (\ref{eqa2}) and (\ref{eqb2}) into (\ref{eqd2}), we
have:
\begin{equation}\label{eqe2}
x_{n-1}=\frac{1}{\lambda_2}(x_{n-1}+y_{n-1}-s_{n-1})+\beta_{n},
\end{equation}
and substitute (\ref{eqc2}) into (\ref{eqe2}), we get:
\begin{equation*}
\beta_n=\frac{1}{\lambda_2}s_{n-1}-\frac{1}{\lambda_2}\beta_{n-1}
\end{equation*}
so we have:
\begin{equation}\label{eqf2}
\beta_n=-\sum_{j\ge 1}(-1)^j\frac{s_{n-j}}{\lambda_2^j}.
\end{equation}

Whenever an initial point $(x_0,y_0)$ is on the intersection of
two straight lines $l_0^-$ and $l_0^+$, then $(x_n,y_n)$ would be
on the intersection of straight lines $l_n^{\pm}$. This process
can continue if the intersection is within $[0,1)^2$, so we have
two of the admissibility conditions for symbolic sequences as
follows
\begin{equation}\label{eq_admis3}
\left\{ \begin{array}{rl} 0\le & \alpha_0-\beta_0=
\sum\limits_{j\le -1}(-1)^j\lambda_2^js_j+
\sum\limits_{j\ge 0}\lambda_2^{-j}s_j < \lambda_2-\lambda_1 \\
0\le & \alpha_0+\lambda_2^2\beta_0= \sum\limits_{j\ge
0}\lambda_2^{-j}s_j- \lambda_2^2\sum\limits_{j\le
-1}(-1)^j\lambda_2^{j}s_j < 1+ \lambda_2^2
\end{array},\right.
\end{equation}

For $(x_0,y_0)\in [0,1)^2$, let its itinerary be
$$
\iota(x_0,y_0)=(\cdots s_{-j}\cdots s_{-1}\cdot s_0s_1\cdots s_j
\cdots).
$$
Suppose
$$
\left\{ \begin{array}{l}
y_0=\frac{1}{\l_1}x_0+K_1\\
y_0=\frac{1}{\l_2}x_0+K_2
\end{array},\right.
$$
then $K_1=y_0-\frac{1}{\l_1}x_0$ and $K_2=y_0-\frac{1}{\l_2}x_0$,
so we should have $\alpha_0=K_1$ and $\beta_0=K_2$, so we have the
conditions for a symbolic sequence to be the itinerary of the
given point $(x_0,y_0)$ as follows
\begin{equation}\label{eq_admis4}
\left\{ \begin{array}{l}
\sum\limits_{j\ge 0}\lambda_2^{-j}s_j=y_0-\frac{1}{\l_1}x_0\\
\sum\limits_{j\le -1}\lambda_2^{j}s_j=\frac{1}{\l_2}x_0-y_0
\end{array}\right.
\end{equation}

{}From the discussion above, the $n$-th iteration $(x_n, y_n)$ is
the intersection of two straight lines $l^-_n$ and $l^+_n$, then
\begin{equation}\label{eq_xnyn}
\left\{ \begin{array}{rl}
x_n= & \frac{\alpha_n-\beta_n}{\lambda_2-\lambda_1}\\
y_n= & \frac{\lambda_2 \beta_n-\lambda_1
\alpha_n}{\lambda_2-\lambda_1}
\end{array}, \ \ n\ge 0 \right.
\end{equation}

By substituting (\ref{eqga}) and (\ref{eqf2}) into
(\ref{eq_xnyn}), we know that there is a one-to-one correspondence
between all points in the phase space (and the orbits from them)
and all admissible symbolic sequences (and their shifts):
$$
(x_n,y_n)\longleftrightarrow (\cdots s_{n-1}\cdot
s_ns_{n+1}\cdots), \ n \ge 0.
$$
\qed

Note that in the case for $a=b=1$, the partition only contains two
elements $P_0=\{(x,y)\in [0,1)^2,x+y<1\}$ and $P_1=\{(x,y)\in
[0,1)^2,x+y\ge 1\}$. So coding $s_n=0$ or $1$.

As applications of Proposition~\ref{propadmiss2}, in the following
we look at fixed points and 2-period points.

For fixed points, all their codings are $0$ or $1$. But
$(\cdots11\cdot 11\cdots)$ is not admissible; for $(\cdots00\cdot
00\cdots)$, $\alpha_n=\beta_n=0, \ -\infty< n< +\infty,$ from
(\ref{eq_xnyn}) we know for all iterations $x_n=y_n=0$, so the
only fixed point is $(0,0)$.

For a $2$-period point $(x^{(2)},y^{(2)})$, the possible
itineraries are:
$$
\iota(x^{(2)},y^{(2)})=\left\{
\begin{array}{l}
{\bf s_0}=(\cdots0101\cdot 0101\cdots)\\
{\bf s_1}=(\cdots1010\cdot 1010\cdots)
\end{array}\right.
$$
therefore
$$
\left\{
\begin{array}{l}
\alpha_0({\bf s_0})=\sum\limits_{j\ge
0}\frac{1}{\lambda_2^{2j+1}}=
\frac{\lambda_2}{\lambda_2^2-1}\\
\alpha_0({\bf s_1})=\sum\limits_{j\ge 0}\frac{1}{\lambda_2^{2j}}=
\frac{\lambda_2^2}{\lambda_2^2-1}  \\
\beta_0({\bf s_0})=-\sum\limits_{j\ge
0}(-1)^{2j+1}\frac{1}{\lambda_2^{2j+1}}=
\frac{\lambda_2}{\lambda_2^2-1} \\
\beta_0({\bf s_1})=-\sum\limits_{j\ge 1}\frac{1}{\lambda_2^{2j}}=
\frac{1}{1-\lambda_2^2}
\end{array}. \right.
$$
Check these with the admissibility conditions (\ref{eq_admis3}),
we find $\alpha_0({\bf s_1})-\beta_0({\bf
s_1})=\lambda_2-\lambda_1$, so ${\bf s_1}$, and therefore ${\bf
s_0}$ as well, are not admissible. This means that there is no
2-period point for the system.

This procedure can be applied to the analysis of periodic points
of higher periods.

\section{Type III linear torus maps}
\label{sec_ellip}

The partition and coding introduced in Section~\ref{sec_linear}
can be applied to all the three types of linear torus maps,
however, from the above discussions, the partition and coding are
more helpful to type I and II maps, as these linear systems on the
2-torus can be viewed as 1-dimensional systems in the sense that
they keep eigen-directions invariant. In
Section~\ref{sec_hyper_para} by using this property we have given
symbolic descriptions about their dynamics. In this section we
will introduce another coding which is more helpful to describe
type III maps.

For the area-preserving cases $\det M =1$, as introduced in
\cite{ADF01} and \cite{AshwinFu01}, the Sigma-Delta modulator map,
which is a piecewise linear elliptic system on the plane, and the
overflow oscillation map, which is an elliptic linear system on
the 2-torus, are equivalent to piecewise rotations, which can be
viewed as a complex system with one variable, by appropriate
transformations of the linearized parts into Jordan normal form.

In general, a type III linear system on the 2-torus
\begin{equation*}
\left\{\begin{array}{l}
x'=ax+by ~(\bmod ~1)\\
y'=cx+dy ~(\bmod ~1)
\end{array}\right.
\end{equation*}
can be viewed as a 1-dimensional complex system in another sense,
i.e., the system can be transformed to complex system with one
variable by linear shear. That is, let $N$ be the matrix such that
$$
N^{-1}(a,b;c,d)N =\sqrt{\Delta} (\cos\theta,-\sin\theta;
\sin\theta,\cos\theta),
$$
where $\Delta= \det M \  (>0)$ is the determinant, and
$\sqrt{\Delta}{\mbox{e}}^{\pm i\theta}$ are two eigenvalues, then
let
$$
\vctr{x}{y}=N\vctr{u}{v}, \ \mbox{ and } z=u+iv,
$$
the linear system on the 2-torus becomes the complex system:
\begin{equation}\label{eq_complex}
z'=\sqrt{\Delta} {\mbox{e}}^{i\theta}z+W_j,
\end{equation}
where $z\in P'_j\subset X, \ X=\bigcup P'_j, \ P'_j$ are
parallelogram partition elements obtained by shearing the original
partition elements $P_{s,t}$ (See figure~\ref{fig_parti}), and
$W_j\in \mathbb C$ are translation terms induced by the
transformation.

For the type of system (\ref{eq_complex}), we can introduce a
coding underlying map operations, which generalizes the one
introduced for the overflow oscillation system in
\cite{AshwinFu01}, and give a symbolic description about the
periodic points. We will present some details in the next section
for planar piecewise isometries.

\subsection{Planar piecewise isometries}
\label{sec_pwi}

{}From (\ref{eq_complex}) a type III torus map with unit
determinant ($\det M =1$) can be represented in Jordan normal form
as a anti-clockwise rotation by $\theta$ followed by a
translation, that is, the map is equivalent to a planar piecewise
isometry after a linear shearing.

In general, we consider planar piecewise isometries
$f:X\rightarrow X$ defined on a partition $\{P_i\}_{i=1}^n$ into
(possibly unbounded) open, convex polygonal atoms, such that
$$
X=\cup_i \overline{P}_i \mbox{ with } P_i\cap P_j=\emptyset \mbox{
for }i\not= j,
$$
and $f$ restricted to each $P_i$ is an isometry. We suppose here
that $f$ is a oriented piecewise isometry, and has a common
rotation angle $\theta$, and the family of PWIs can be
parameterized by the $\theta$. So $f$ can be written as
\begin{equation}\label{eq_map_symb}
f(z)= \om z + W_j, \ \mbox{ if } z\in P_j, \ j=0,1,\cdots,N-1.
\end{equation}
Where $\om = {\mbox{e}}^{i\theta}$. For rational $\theta/\pi$,
usually $f$ either perform a periodic permutation or polygon
packing where all points in the polygons are periodic. We will be
interested in the cases where $\theta/\pi$ is irrational. i.e.,
$f$ is an irrational rearrangement.

By using the coding introduced in Appendix~B of \cite{Futhesis},
for a periodic point $z_{\bf k}$ of period $n$ with itinerary
$\iota(z_{\bf k})=P(k_0k_1\cdots k_{n-1})$, where
$$
k_j=(k^{(0)}_j,k^{(1)}_j,\cdots,k^{(N-1)}_j)\in
\{e_0,e_1,\cdots,e_{N-1}\}, \ e_j=(\delta_{0j},\delta_{1j},
\cdots, \delta_{(N-1)j}),
$$
we have
\begin{equation}\label{eq_gnrlzk_symb}
z_{\bf k}=\frac{1}{1-\om ^n}\sum_{j=0}^{n-1}\om ^j
\sum_{s=0}^{N-1}W_sk^{(s)}_{n-1-j}.
\end{equation}
So we have the following result (there is a version of the
result below in \cite{Goetz00b}, here we give a new proof
by using (\ref{eq_gnrlzk_symb})):

\begin{proposition}
There is a one-to-one correspondence between admissible periodic
symbolic sequences and periodic points of $f$ as long as
$\frac{\theta}{\pi} \notin \mathbb Q$.
\end{proposition}

\proof We can use (\ref{eq_gnrlzk_symb}) to prove this. Let
$$
\iota(z_{\bf k}) =\iota(z'_{\bf k}) =P(k_0k_1\cdots k_{n-1}),
$$
then from (\ref{eq_gnrlzk_symb}) we have
$$
(1-\om^n)(z_{\bf k}-z'_{\bf k})=0,
$$
but $1-\om^n\ne 0$ since $\theta/\pi\notin \mathbb Q$, so it must
have $z_{\bf k}=z'_{\bf k}$. \qed

In \cite{AshwinFu01}, \cite{AFD01} and Appendix~B of
\cite{Futhesis} the explicit expression (\ref{eq_gnrlzk_symb}) is
used to prove that tangencies between discs in the invariant disc
packings for the overflow oscillation, Sigma-Delta modulator, and
some general planar piecewise isometries are very rare. We expect
this can be applied to other problems for PWIs.

By the way, a piecewise isometric system can have attractors although
its all Lyapunov exponents are zero (\cite{AshwinFu02,mends}).

\section{Some remarks}
\label{sec_final}

\subsection{Linear torus maps in higher dimensions}

The techniques and ideas used in this paper may be generalized to
higher dimensional piecewise linear maps, e.g., the similar maps
on the $m$-torus, while the classification may be more complicated
than the 2-torus case, as we have to deal with matrices with mixed
real and complex eigenvalues, leading to some kind of mixed types.

For the following class of maps $({x'}^{(1)},\cdots,
{x'}^{(m)})=f(x^{(1)},\cdots, x^{(m)})$ of the $m$-torus
$X=[0,1]^m$ of the form
\begin{equation}\label{eq_m-torusmap}
\left\{\begin{array}{rcll}
{x'}^{(1)} & = & a_{11}x^{(1)}+\cdots+a_{1m}x^{(m)} & ~(\bmod ~1)\\
         & \vdots &  & \\
{x'}^{(m)} & = & a_{m1}x^{(1)}+\cdots+a_{mm}x^{(m)}  & ~(\bmod ~1)
\end{array}\right.
\end{equation}
at first we suppose
$M=(a_{11},\cdots,a_{1m};\cdots;a_{m1},\cdots,a_{mm})$ has a real
eigenvalue $\lambda$, and the $n$-th iteration $(x^{(1)}_n,
\cdots, x^{(m)}_n)$ of the map with coding $(s^{(1)}_n, \cdots,
s^{(m)}_n)$ is located on the hyperplane $l_n$:
$$
x^{(m)}=\alpha_1x^{(1)}+\cdots+\alpha_{m-1}x^{(m-1)}+\delta_n,
$$
where $s^{(j)}_n =\lfloor
a_{j1}x^{(1)}+\cdots+a_{jm}x^{(m)}\rfloor, n\ge 0$, and
$(\alpha_1,\cdots, \alpha_{m-1}, -1)$ is an eigenvector for the
transposed matrix $M^T$ with respect to the eigenvalue $\lambda$.
Then similar to the discussion in Section~\ref{sec_hyper_para} we
have:
\begin{equation}\label{eq_delta}
\delta_{n+1}=\sum_{i=1}^{m}\alpha_is^{(i)}_n+\lambda \delta_n, \
n\ge 0
\end{equation}
where $\alpha_m=-1$.

That is, the map (\ref{eq_m-torusmap}) sends points on hyperplane
$l_n$ to points on hyperplane $l_{n+1}$, and hyperplane $l_{n+1}$
can be determined through (\ref{eq_delta}) by $l_n$ and coding
$(s^{(1)}_n, \cdots, s^{(m)}_n)$. In this sense, the
$m$-dimensional system (\ref{eq_m-torusmap}) possesses
$(m-1)$-dimensional dynamics.

(\ref{eq_delta}) can be rewritten as
\begin{equation*}
\delta_n=-\frac{1}{\lambda}\sum_{i=1}^{m}\alpha_is^{(i)}_n
+\frac{1}{\lambda}\delta_{n+1},
\end{equation*}
so we have the relation among $\delta_j$ and the codings
$(s^{(1)}_j, \cdots, s^{(m)}_j)$ for $0\le n\le j \le n+J+1$:
\begin{equation}\label{eq_delta2}
\delta_n=-\sum_{j=0}^{J}\frac{1}{\lambda^{j+1}}
\sum_{i=1}^{m}\alpha_is^{(i)}_{n+j}+\frac{1}{\lambda^{J+1}}\delta_{n+J+1}
\end{equation}
Under some conditions, say $|\lambda|>1$, we have:
\begin{equation}\label{eq_delta_series}
\delta_n=-\sum_{j=0}^{+\infty}\frac{1}{\lambda^{j+1}}
\sum_{i=1}^{m}\alpha_is^{(i)}_{n+j}
\end{equation}

Now we suppose $m=2m'$ be even and all eigenvalues are complex,
and suppose that there exists a matrix $N$ such that
$$
N^{-1}MN=\Delta^{\frac{1}{m}}\mbox{diag}(M_1,\cdots,M_{m'}),
$$
where $M_k=(\cos \theta_k,-\sin \theta_k;\sin \theta_k,\cos
\theta_k)$. Let
$$
\left(\begin{array}{c}
x^{(1)}\\
\vdots\\
x^{(m)}
\end{array}
\right) =N \left(\begin{array}{c}
u^{(1)}\\
v^{(1)}\\
\vdots\\
u^{(m')}\\
v^{(m')}
\end{array}
\right), \ \mbox{ and } z^{(k)}=u^{(k)}+i v^{(k)}, \
k=1,\cdots,m',
$$
then similar to the $2$-dimensional case, the linear system
(\ref{eq_m-torusmap}) on the $m$-torus can be transformed to a
complex system with $\frac{m}{2}$ variables, which is a rotation
on each component.

\subsection{Admissibility conditions for planar PWIs}

For the elliptic cases, admissibility conditions for periodic
itineraries are crucial in estimating the measure of the
exceptional sets. To see what are the situations we may face, here
as an example we remark briefly on the piecewise isometry of the
torus obtained from the overflow oscillation problem (see
\cite{AshwinFu02} for details).

The piecewise isometry $f:M_\theta\rightarrow M_\theta$ is defined
by
$$
f(z)= \omega z +W k(z),
$$
where $\theta\in(0,\pi/2)$, $M_\theta$ is the rhombus unit cell
for a torus:
$$
M_\theta=\{ z\in\C~:~ |\re(z)|\leq 1 ~\mbox{ and }~ |\re(z
e^{i\theta})|\leq 1\}
$$
and $\omega = {\rm e}^{i \theta}, W=-\frac{2i}{\sin \theta}$ is a
constant, and
$$
k(z)=\left\{\begin{array}{rl}
-1 & \mbox{ if } z\in M_{-1} \\
+1 & \mbox{ if } z\in M_{+1}\\
0 & \mbox{ if } z\in M_0
\end{array}\right.
$$
is the coding, where
$$
\begin{array}{ccl}
M_{-1} &=& \{z\in M_\theta, \re(z e^{2 i\theta})>1 \},\\
M_{+1} &=& \{z\in M_\theta, \re(z e^{2 i\theta})\leq -1 \},\\
   M_0 &=& M_\theta\setminus (M_{-1}\bigcup M_{+1}).
\end{array}
$$
For a point $z_0$ with itinerary ${\bf k}=(k_0k_1\cdots
k_j\cdots)$, we have
$$
z_{n+1}=\omega z_n+Wk_n, \ n\ge 0.
$$
Therefore we have
\begin{eqnarray*}
z_n & = & -W\frac{k_n}{\omega}+\frac{z_{n+1}}{\omega}\\
    & = & -W\sum_{j=0}^{J}\frac{k_{n+j}}{\omega^{j+1}}
         +\frac{z_{n+J+1}}{\omega^{J+1}}.
\end{eqnarray*}
In form we can write
\begin{eqnarray*}
z_0 & = & -W\sum_{j=0}^{+\infty}\frac{k_j}{\omega^{j+1}}\\
    & = & -W \left(\sum_{j=1}^{+\infty}k_{j-1}\cos j\theta
          -i \sum_{j=1}^{+\infty}k_{j-1}\sin j\theta \right).
\end{eqnarray*}
It is noted that the convergence of the above trigonometric series
is indefinite. The above analysis is applicable to all planar
piecewise isometries. So we can say that this is the elementary
but the main difficulty involved in finding admissibility
conditions for itineraries in the symbolic dynamics analysis for
planar piecewise isometries. Recently Yu and Galias studied
similar problem and obtained some numerical results on admissible
periodic symbolic sequences (\cite{YuGal01}, see also
\cite{GalOgo92,Vow99} for earlier works).

\subsection*{Acknowledgement}

The authors would like to thank the UK Engineering and Physical
Science Research Council for support via grant GR/M36335. XF also
thanks School of Mathematical Sciences of Exeter University for
generous support when the earlier version of this paper was
written.sd

\end{document}